\documentclass[showpacs,preprintnumbers,amsmath,aps,amssymb]{revtex4}
\usepackage{graphicx}% Include figure files
\usepackage[english]{babel}
\usepackage{bm}% bold math
\begin{document}
\baselineskip 16pt
\title{Variable cosmological term $\Lambda(t)$.}% in covariant way}
\author{J. Socorro$^1$}
\email{socorro@fisica.ugto.mx}
\author{M. D'oleire$^2$}
\email{madiaz@stud.ntnu.no}
\author{Luis O. Pimentel$^3$}
\email{lopr@xanum.uam.mx}
 \affiliation{$^1$Departamento de  F\'{\i}sica, DCeI, Universidad
de Guanajuato-Campus Le\'on,
 C.P. 37150, Le\'on, Guanajuato, M\'exico\\
$^2$ Norwegian University of Science and Technology, Department of
Petroleum Engineering and Applied Geophysics Trondheim, Norway\\
  $^3$ Departamento de F\'isica, Universidad Aut\'onoma Metropolitana, Apartado Postal 55-534,
C.P. 09340 M\'exico, DF, M\'exico
}%
\begin{abstract}
We present the case of time-varying cosmological term $\Lambda(t)$.
The main idea arises by proposing that as in the cosmological constant
case, the scalar potential is identified as $ V(\phi)=2\Lambda$,
with $\Lambda$ a constant, this identification should be kept even when
the cosmological term
has a temporal dependence, i.e.,  $ V(\phi(t))=2\Lambda(t)$. We Use
the Lagrangian formalism for a scalar field $\phi$ with
standard  kinetic energy  and arbitrary potential
$V(\phi)$ and apply this model  to the Friedmann-Robertson-Walker (FRW)
cosmology. Exact solutions of the field equations are obtained by  a
special ansatz to solve the Einstein-Klein-Gordon equation and a
particular  potential for the scalar field  and barotropic perfect
fluid. We present the  evolution on this cosmological term with
different scenarios.
\end{abstract}
 \maketitle

\section{introduction}
The present phase of an accelerated expansion of the universe stands
as one of the most challenging open problems in modern cosmology and
astrophysics. This acceleration is characterized by
 which is popularly known as dark
energy. Among many possible alternatives, the simplest candidate for
dark energy is the vacuum energy which is mathematically equivalent
to the cosmological constant. Models with different decay laws for
the variation of cosmological term were investigated during the last
two decades in a non covariant way, (Chen \& Wu 1990); (Abdel 1990);
(Pavon 1991); (Carvalho et al 1992); (Kalligas et al 1992); (Lima \&
Maia 1994);  (Lima \& Carvalho 1994); (Lima \& Trodden 1996); (Arbab
\&  Abdel 1994); (Birkel \& Sarkar 1997); (Silveira \& Waga 1997);
(Starobinsky 1998); (Overduin \& Cooperstock 1998); (Vishwakarma
2000,2001); (Arbab 2001,2003,2004); (Cunha \& Santos 2004);
(Carneiro \& Lima 2005); (Fomin et al 2005); (Sola \& Stefancic
2005,2006); (Pradhan et al 2007); (Jamil \& Debnath 2011) and
(Mukhopadhyay 2011); in particular, in Fomin et al (2005) there are
several evolution relations for $\Lambda$ which many author have
used, also in Ref. Overduin \& Cooperstock (1998) appears a table
with these relations and the corresponding references where they
were considered. Anisotropic cosmological models, also has been
treated in this formalism from different points of view (Aroonkumar
1993,1994); (Arbab 1997); (Singh et al 1998); (Pradhan \& Kumar
2001); (Pradhan 2003,2007,2009); (Pradhan \& Pandey 2003,2006);
(Pradhan et al 2007,2008); (Carneiro 2005); (Esposito et al 2007);
(Bal \& Singh 2008); (Belinch\'on 2008); (Singh et al 2008), (Singh
et al 2013); (Shen 2013); (Tripathy 2013) and (Rahman \& Ansary
2013).

In this work we present an analysis in covariant way, using the
Lagrangian density of standard scalar field. The main idea arises by
proposing that as  in the cosmological constant case, the scalar
potential is identified as $ V(\phi)=2\Lambda$, with $\Lambda$ a
constant. So, we {\it extend} this idea and suggest that this
correspondence is valid even when this cosmological term has a
temporal dependence, i.e.,  $ V(\phi(t))=2\Lambda(t)$. We include a
barotropic equation state between the pressure and energy density of
the scalar field, $\rm p_\phi= \omega_\phi \rho_\phi$, quantities
that we shall define in the following lines. In order to built up
the analysis presented here, initially we solve the Klein-Gordon
equations, whose solution implies that the energy density of a
scalar field has a wide range of scaling behavior, $\rho_\phi \sim
A^{-m}$ with A the scale factor of the FRW model, (Ferreira \& Joyce
1998); (Liddle \& Sharrer 1998) and (Copeland et all 1998), that
emerges as a proportionality law between the energy density of the
scalar field and the energy density of the barotropic perfect fluid,
relation that is know as an "attractor solution", with the
proportionality constant $\rm m_\phi$ (Liddle \& Sharrer 1998), that
is, $\rm \rho_\phi=m_\phi \rho$.  However the nature of the matter
that corresponds to the scalar field is unknown since  this matter
is not detectable by the usual methods.

The research in  non covariant way goes  by
beginning with the relation when the cosmological term $ \Lambda$, often  treated as a constant, having a geometrical
interpretation. For explain this, the Einstein field equation is
written as
\begin{equation}
\rm G_{\mu \nu} + \Lambda g_{\mu \nu}=-8\pi G T_{\mu \nu},
\label{EFE}
\end{equation}
where $\rm G_{\mu \nu}$ is the usual Einstein tensor and $\rm T_{\mu
\nu}$ is the energy-momentum tensor of matter. When we take the
covariant divergence of this equation, the vanishing of the Einstein tensor  is
guaranteed by the Bianchi identities, then it is assumed that the
energy-momentum tensor satisfies the corresponding conservation law
$\rm \nabla^\nu T_{\mu \nu}=0$, and that the covariant divergence of
cosmological term must vanish, this implies that $\rm
\Lambda=constant$. Usually, this argument situates this cosmological
constant on the left-hand side of the field equations, given a
geometrical interpretation of the cosmological term.

However, if the cosmological term is moved to  the right-hand side
of the field equation, the interpretation of $\rm \Lambda$ change as
part of the matter content in the following sense. The field
equations now are written as
\begin{equation}
\rm G_{\mu \nu} =-8\pi G \tilde T_{\mu \nu}, \qquad  \tilde T_{\mu
\nu}= T_{\mu \nu}+ \frac{\Lambda}{8\pi G} g_{\mu \nu}. \label{EFE-N}
\end{equation}
Once this is done, there is  no a priori reason why this
cosmological term should not vary, considering that it is  the effective
energy-momentum tensor $\rm \tilde T_{\mu \nu}$ that satisfies the
conservation law
\begin{equation} \rm \nabla^\nu \tilde T_{\mu \nu}=0,\label{C-l}
\end{equation}.

The set of equation (\ref{EFE-N}, \ref{C-l}) and one state equation,
are the fundamental tools to do the research, considering a variable
cosmological term in non covariant way, because there is no a
Lagrangian density that reproduce these field equations
(\ref{EFE-N}, \ref{C-l}).

In the present treatment we take into  account the corresponding
Lagrangian density with a scalar field
\begin{equation}
 \rm {\cal L}[g,\phi]=\sqrt{-g}\left(R-\frac{1}{2}g^{\mu \nu}\nabla_\mu\phi\nabla_\nu\phi
 +V(\phi)\right)+\sqrt{-g} {\cal L}_{matter}\label{lagra}
\end{equation}
where R is the Ricci scalar, ${\cal L}_{matter}$ correspond to a
barotropic perfect fluid, $\rm p=\omega \rho$, $\rho$ is the energy
density and p is the pressure of the fluid in co-moving frame and
$\omega$ is the barotropic constant.

The corresponding variation of  (\ref{lagra}), with respect to the metric and
the scalar field gives the Einstein  and Klein-Gordon field equations
\begin{eqnarray}
&&\rm
R_{\alpha\beta}-\frac{1}{2}g_{\alpha\beta}R=-\frac{1}{2}\left(\nabla_\alpha\phi\nabla_\beta\phi-\frac{1}{2}g_{\alpha\beta}
g_{\mu\nu}\nabla_\mu\phi\nabla_\nu\phi\right)+\frac{1}{2}g_{\alpha\beta}V(\phi)-8\pi GT_{\alpha\beta}, \label{camrel}\\
&&\rm \square\phi-\frac{\partial V}{\partial\phi}=0.\label{klein}
\end{eqnarray}
From  (\ref{camrel}) it  can be deduced  that the energy-momentum
tensor associated with the scalar field is
\begin{equation}
\rm
8 \pi G T^{(\phi)}_{\alpha\beta}=\frac{1}{2}\left(\nabla_\alpha\phi\nabla_\beta\phi-\frac{1}{2}g_{\alpha\beta}
g_{\mu\nu}\nabla_\mu\phi\nabla_\nu\phi\right)-\frac{1}{2}g_{\alpha\beta}V(\phi)
\end{equation}\\
and the corresponding tensor for a barotropic perfect fluid  becomes
$$\rm T_{\alpha \beta}= \left(p+\rho\right) u_\alpha u_\beta +
g_{\alpha \beta} p$$
here $\rm u_\alpha$ is the four-velocity in the
comoving  frame. The line element to be considered in this work is the FRW
\begin{equation}
\rm ds^2=-N(t)^2 dt^2 +A(t)^2 \left[\frac{dr^2}{1-\kappa r^2}
+r^2(d\theta^2+sin^2\theta d\phi^2) \right]=-d\tau^2 + A(\tau)^2
\left[\frac{dr^2}{1-\kappa r^2} +r^2(d\theta^2+sin^2\theta d\phi^2)
\right], \label{frw}
\end{equation}
where we identify the time transformation $\rm N(t) dt=d\tau$, this
transformation will be used in the whole work, and in special gauge
we recover directly the cosmic time t.
 \section{field equations}
 Making use to the metric (\ref{frw}) and a comoving  fluid, the equations (\ref{camrel}) y (\ref{klein})
 becomes (a dot mean a time derivative)
\begin{eqnarray}\rm
\frac{3\dot{A}^2}{A^2}+\frac{3\kappa N^2}{A^2}-8\pi G\rho N^2-\frac{1}{4}\dot{\phi}^2-\frac{N^2V(\phi)}{2}&=&0,\label{ein0}\\
\rm
\frac{2\ddot{A}A}{N^2}+\frac{\dot{A}^2}{N^2}-\frac{2\dot{A}\dot{N}A}{N^3}+\kappa+8\pi
G A^2p+\frac{ A^2\dot{\phi}^2}{4N^2}
-\frac{1}{2}A^2V(\phi)&=&0,\label{eini}\\
\rm \left[-3\frac{\dot A}{A} \frac{{\dot \phi}}{N^2}-\frac{\ddot
\phi}{N^2}+\frac{\dot \phi}{N}\frac{\dot
N}{N^2}\right]-\frac{\partial V}{\partial \phi}&=&0,
\end{eqnarray}
using the time transformation $\rm d\tau=Ndt$, and the chain rule
$\frac{\partial V}{\partial\phi}=\frac{\partial V}{\partial
\tau}\frac{\partial \tau}{\partial\phi}=
\frac{V^\prime}{\phi^\prime}$ in the last equation we obtain

\begin{eqnarray}\rm \frac{3A^{\prime 2}}{A^2}+\frac{3\kappa }{A^2}-8\pi G\rho -\frac{1}{4}\phi^{\prime 2}-\frac{V(\phi)}{2}&=&0, \label{co0} \\
\rm \frac{2A^{\prime \prime}}{A}+\frac{A^{\prime
2}}{A^2}+\frac{\kappa}{A^2}+8\pi G p+\frac{1}{4} \phi^{\prime
2}-\frac{1}{2}V(\phi)&=&0, \label{co1}\\
 \left[3\frac{ A^\prime}{A} \phi^{\prime2}+ \phi^\prime\phi^{\prime\prime}\right]&=&\rm -V^\prime\label{kgord},
\end{eqnarray}
the Klein-Gordon equation (\ref{kgord}) can be rewritten as
\begin{equation}
\rm \frac{d}{d\tau}\left[Ln\left(A^6 \frac{\phi^{\prime
2}}{2}\right) \right]=-\frac{V^\prime}{\frac{\phi^{\prime
2}}{2}},\label{kg0}
\end{equation}
In the literature there are  some articles where the authors  try to
solve these field equation in general way, for instance, in
reference Chimento \& Jakubi (1996), the authors present an
elaborate technique for solve the Klein-Gordon equation
(\ref{kgord}), and in
 Reyes (2008) they  use an algebraic method to obtain exact
solutions taking as the basic variable the energy density of the
scalar field.

In order to solve this set of equations, we introduce the ansatz, of
considering  that the energy density of the field $\phi$ is
proportional to the  energy density of the  barotropic perfect
fluid, $\rho_\phi= m_\phi \rho$, where $\rm m_\phi$ is a {\it
positive constant}. The scaling behavior occurs when $\rm m_\phi<1$,
otherwise, the quintessence field is dominant.

 The energy density and pressure of the field $\phi$ are
given as
 $$\rm 16\pi G\rho_\phi=\frac{1}{2}\phi^{\prime 2}+ V(\phi), \qquad
16\pi Gp_\phi= \frac{1}{2}\phi^{\prime 2}- V(\phi)$$
now, equations (\ref{co0},\ref{co1}) are rewritten as
\begin{eqnarray}
\rm \frac{3A^{\prime 2}}{A^2}+\frac{3\kappa }{A^2}-8\pi G\left(\rho +\rho_\phi\right)&=&0, \label{co00} \\
\rm \frac{2A^{\prime \prime}}{A}+\frac{A^{\prime
2}}{A^2}+\frac{\kappa}{A^2}+8\pi G\left(p+ p_\phi\right)&=&0.
\label{co11}
\end{eqnarray}
We will make now the assumption that the scalar is a  barotropic
fluid: $\rm p_\phi= \omega_\phi \rho_\phi$, where $\rm \omega_\phi$
is a constant that play the same role of the $\omega$ parameter in
the barotropic perfect fluid. Under this proposal, the field
equations are
\begin{eqnarray}
\rm \frac{3A^{\prime 2}}{A^2}+\frac{3\kappa }{A^2}-8\pi G\rho_T&=&0, \label{co000} \\
\rm \frac{2A^{\prime \prime}}{A}+\frac{A^{\prime 2}}{A^2}+\frac{\kappa}{A^2}+8\pi G p_T
&=&0, \label{co111}\\
\rm \frac{d}{d\tau}\left[Ln\left(A^6 \frac{\phi^{\prime 2}}{2}
\right) \right]&=&\rm -\frac{V^\prime}{\frac{\phi^{\prime 2}}{2}},
\label{kg000}
\end{eqnarray}
where  the total energy density is  $\rho_T=\rho+ \rho_\phi=\alpha_\phi \rho$,
with $\rm \alpha_\phi=1+m_\phi>1$, and the total  pressure is
$p_T=p+p_\phi=\omega_T \rho_T$, with the last barotropic index is
given by
\begin{equation}
\rm \omega_T=\frac{\omega+m_\phi \omega_\phi}{\alpha_\phi}
\label{index-T}.
\end{equation}

Equation (\ref{kg000}) can be written as
\begin{equation} \rm
\frac{d}{d\tau}\left[Ln\left(A^6
V^{\frac{2}{1+\omega_\phi}}\right) \right]=0, %\label{pot-form}
\end{equation}
whose solution is
\begin{equation} \rm
\qquad V(\tau)=c_\omega  \,A^{-3(1+\omega_\phi)} \label{VA}
\end{equation}
and the relation  $\rm p_\phi=\omega_\phi \rho_\phi$ implies  that
$\rho_\phi=\frac{2 c_\omega}{1-\omega_\phi}
A^{-3(1+\omega_\phi)}\sim A^{-m}$ with $\rm m=3(1+\omega_\phi)$ and
considering the solution to the energy-momentum tensor of the
perfect fluid $\rm \nabla_\nu T^{\mu \nu}=0$, with solution as $\rm
\rho=M_\omega A^{-3(1+\omega)}\sim A^{-n}$, with $\rm
n=3(1+\omega)$, where initially the two barotropic indexes $\rm
\omega_\phi$ and $\omega$ are different. Is know in the literature
that the case $\rm m=n$ gives an ''attractor solution'' and
corresponds to the case when the potential of the scalar field
$\phi$ goes to exponential behavior; this case has been studied,
using other methods to understand the evolution of the universe,
where this potential is introduced by hand (Lucchin \& Matarrese
1985); (Halliwell 1985); (Burd \& Barrow 1988); (Wetterich
1998);(Wand et al 1993); (Ferreira \& Joyce 1997) and (Copeland et
all 1998).

In order to solve this set of equations, we consider the  case $\rm
m=n$, thus we have that $\omega=\omega_\phi$ implying that $\rm
\omega_T=\omega=\omega_\phi$ and then  find the corresponding potential of
the scalar field for a wide range of values of the barotropic index
$\omega_\phi$.

This last result can be obtained making the following analysis,
considering that we  have three barotropic indices, two assumed, $
\omega$ and $\omega_{\phi}$, and one implied, $\omega_T$. We show
now that all three must be the same. Notice that he first two
equations are the standard FRW equation for the total fluid with
barotropic index $\omega_T$.  Also notice that under the assumption
of the relation between  the density of the scalar field and the
density of the perfect fluid, and the proportionality of the
$p_\phi$ with $\rho_\phi$ then  the potential V, ${\phi'}^2 /2$  and
all  the densities ($\rho, \rho_\phi,\rho_T$)are proportional to
each other. From Eqs. (\ref{co000},\ref{co111}) we now that
\begin{equation} \rm
\rho_T = c_T A^{-3(\omega_T+1)}.
\label{rhoTA}
\end{equation}
Then as a consequence of the proportionality between $V $ and
$\rho_T$ the exponents in Eqs. (\ref{rhoTA}, \ref{VA}) should be
equal, that corresponds to $\rm m=n$ case,
\begin{equation} \rm
 \omega_T =  \omega_\phi,\qquad \Rightarrow
\qquad    \omega= \omega_\phi=\omega_T.
\end{equation}

\section{General solution for flat space}

In this section we present solutions to the field equations for the flat case.
Equation (\ref{kg000}) is in the flat case written as
\begin{equation} \rm
\frac{d}{d\tau}\left[Ln\left(A^6
V^{\frac{2}{1+\omega}}\right) \right]=0,\qquad \Rightarrow
\qquad V(\tau)=c_\omega  \,A^{-3(1+\omega)}, \label{pot-form}
\end{equation}
using the well known time evolution of the scalar factor for the
barotropic fluid,reported in different places, in particular in
reference Berbena et al (2007) that for future convenience  we write
as
\begin{equation}\label{scale-factor}
 \rm A_\omega(\tau)=\left\{
 \begin{tabular}{ll}
 $\rm \left[a_\omega \, \tau
\right]^{\frac{2}{3(\omega+1)}}, \qquad
a_\omega=(\omega+1)\sqrt{6\pi G \alpha_\phi M_\omega} ,$ &$\rm \quad
\omega\not= -1$ \\
$\rm e^{2\sqrt{\frac{2}{3}\pi G \alpha_\phi M_{-1}}\tau}$&$\rm \quad
\omega=-1$
\end{tabular}
\right.
\end{equation}
The Hubble function $\rm H=\frac{\dot A}{A}=N\frac{A^\prime}{A}$ and
the deceleration parameter
\begin{equation}
\rm q=-\frac{\ddot A \, A}{\dot A^2}=-\frac{\dot H +
H^2}{H^2}=-\left[\frac{A^{\prime \prime} \, A}{A^{\prime 2}} +
\frac{A}{A^\prime}\frac{ N^\prime}{N } \right], \label{deceleration}
\end{equation}
 can
be calculated for our model, and are depending of the gauge shift
function N, which should be important employing the observational
data of Supernova type Ia, see references (Riess et al 1998) and
(Perlmutter et al 1999).

In the gauge N=1, the Hubble function become
\begin{equation} \rm H_\omega(t)=\left\{
\begin{tabular}{ll}
 $\rm \frac{2}{3(\omega+1)}\frac{1}{t},
$ &$\rm \quad
\omega\not= -1$ \\
$\rm 2\sqrt{\frac{2}{3}\pi G \alpha_\phi M_{-1}}$&$\rm \quad
\omega=-1$
\end{tabular}
 \right.
\end{equation}
and the corresponding deceleration parameter
\begin{equation}
 \rm q_\omega=\left\{
\begin{tabular}{ll}
 $\rm \frac{1+3\omega}{2},
$ &$\rm \quad
\omega\not= -1$ \\
$-1$&$\rm \quad \omega=-1$
\end{tabular}
 \right.
\end{equation}
in this gauge, the deceleration parameter have a positive value for
all values $\omega\not= -1$, and only in the exponential behavior
have a negative value.

These results are in agreement with the solution at the deceleration
parameter (\ref{deceleration}) when we take the election that this
one is time dependent, (Pradhan et al 2012), or in more general
sense, we can choose that $\rm q(H)=F(H)$, where H is the Hubble
function. By example, when we choose that $\rm F(H)=-1$, we recover
that the scale factor have a exponential behavior, or $\rm
F(H)=\ell=constant>0$, we can recover the power law in the scale
factor.

Employing the gauge $\rm N \to A^3$, we obtain the following
\begin{equation} \rm H_\omega(t)=\left\{
\begin{tabular}{ll}
 $\rm \frac{2}{3(\omega+1)} \left[a_\omega\right]^{\frac{2}{1+\omega}} \tau^{\frac{1-\omega}{1+\omega}},
$ &$\rm \quad
\omega\not= -1$ \\
$\rm 2\sqrt{\frac{2}{3}\pi G \alpha_\phi M_{-1}}
e^{6\sqrt{\frac{2}{3}\pi G \alpha_\phi M_{-1}}\tau}$&$\rm \quad
\omega=-1$
\end{tabular}
 \right.
\end{equation}
and the corresponding deceleration parameter
\begin{equation}
 \rm q_\omega=\left\{
\begin{tabular}{ll}
 $\rm -\frac{5-3\omega}{2}<0,
$ &$\rm \quad
\omega\not= -1$ \\
$-4$&$\rm \quad \omega=-1$
\end{tabular}
 \right.
\end{equation}
Over these last results, is worthy to mention that is necessary to
choose an appropriate gauge for obtain a negative deceleration
parameter in the transformed time $\tau$.

\subsection{Case $\omega\not=\pm 1$ }
The cases $\omega=\pm 1$ will be considered below, in separate way.

Following with our analysis, the temporal dependence of the
potential becomes
\begin{equation}
\rm V(\tau)=c_\omega \frac{1}{(a_\omega\tau)^{2}},\qquad \Rightarrow
\qquad \Delta \phi=\ell_\omega \, Ln(\tau),
\end{equation}
 where the constants value $\rm c_\omega$ and $\ell_\omega$ are
 determined after substitution into the complete set of  Einstein equations, with the scale factor solution
(\ref{scale-factor}), being the general solutions for any $\omega
\not=\, \pm 1$ the following relations are obtained
\begin{eqnarray}
\rm V_\omega(\tau)&=&\rm
\frac{2m_\phi(1-\omega)}{3(1+\omega)^2\,\alpha_\phi}\,\frac{1}{\tau^2},\qquad
\Leftrightarrow \qquad \Lambda(\tau)=
\frac{m_\phi(1-\omega)}{3(1+\omega)^2 \alpha_\phi}\,\frac{1}{\tau^2},\label{potencial}\\
\rm \Delta \phi(\tau)&=&\rm
\sqrt{\frac{4m_\phi}{3(1+\omega)\alpha_\phi}}\,
Ln(\tau), \label{kinetic}
\end{eqnarray}
when the time $\tau$ is eliminated between the two last equations we obtain the potential (or $\Lambda$) as a function of the scalar field
\begin{equation}
\rm V(\phi)=\rm
\frac{2m_\phi(1-\omega)}{3(1+\omega)^2\alpha_\phi}\,
  e^{-\sqrt{\frac{3(1+\omega)(1+m_\phi)}{{m_\phi}}}\Delta \phi}  , \quad
\Leftrightarrow \quad
\Lambda(\phi)=\frac{m_\phi(1-\omega)}{3(1+\omega)^2\alpha_\phi}\,
    e^{-\sqrt{\frac{3(1+\omega)(1+m_\phi)}{{m_\phi}}}\Delta \phi} .
\label{pot-phi}
\end{equation}
The corresponding scalar potential that emerges from the temporal
solution has, as is cited in the literature,  an exponential
behavior, (Lucchin \& Matarrese 1985); (Halliwell 1985); (Burd \&
Barrow 1988); (Wetterich 1998);(Wand et al 1993); (Ferreira \& Joyce
1997) and (Copeland et all 1998). We observe here that for all
values of the barotropic parameter we have a decreasing cosmological
function in time (remember the relation between the potential energy
and the cosmological term, $\rm V(\tau)=2\Lambda(\tau)$), and that
as function of the scalar field we have an exponential.

\subsection{case $\omega=-1$.}
In this case, the barotropic equation of state implies that the scalar field is constant
as is the potential and we are back to
$\Lambda=constant$ case, also the total energy density is  constant and we have  the exponential expansion factor of equation (\ref{scale-factor}).

\subsection{Case $\omega=1$}
In this case, we obtain in our proposal that $\rm V(\phi)=0$ and
then we have a free scalar field that is  the simplest case of the
k-essence theory or Sa\'ez-Ballester, see reference Sabido et al
(2010) and references therein for complete solutions for FRW
cosmological model and Socorro et al (2014) for the corresponding
Bianchi type I anisotropic cosmological model.

In what follows we consider particular cases, divided in  two
branches of the barotropic parameter  $\rm 0<\omega<1$, and  $\rm
-1<\omega<0$.
\section{Particular solutions}
In this section we consider particular cases of the solutions with specific values of the barotropic
coefficient that are of interest in cosmology and considering positive and negative branches.
\subsection{Positive branch:  $\rm 0\le \omega<1.$}
The relation between the kinetic term and the potential energy of the scalar field is
\begin{equation}
\frac{1}{2}\phi^{\prime 2}=\frac{1+\omega}{1-\omega}V(\phi)
\end{equation}
and we have the following particular case for the barotropic
parameter that are of interest in cosmology and astrophysics.
\begin{enumerate}
\item{} Dust scenario, $\omega=0$.

The set of equations (\ref{potencial},\ref{kinetic},\ref{pot-phi})
have the following form
\begin{eqnarray}
\rm V_\omega(\tau)&=&\rm \frac{2
m_\phi}{3\alpha_\phi}\frac{1}{\tau^2},
\label{dust}\\
\rm \Delta \phi(\tau)&=&\rm \sqrt{\frac{4m_\phi}{3\alpha_\phi}}
\,Ln(\tau).
\label{phi-dust}\\
\rm V(\phi)&=&\rm \frac{2m_\phi}{3\alpha_\phi}\,
e^{-\sqrt{3\left(1+\frac{1}{m_\phi}\right)}\Delta \phi}. \label{pot}
\end{eqnarray}
together with  the following scale
factor
\begin{equation} \rm A(\tau)=\left[a_0 \, \tau
\right]^{\frac{2}{3}}, a_0=\sqrt{6\pi G \alpha_\phi M_0}
%\label{dust}
.
\end{equation}

Also this behavior is found using dynamical system and fitting that
one critical point will be an attractor, obtaining that the
corresponding factor $\lambda$ in the exponential function $\rm
V(\phi)\approx e^{\lambda \phi}$, will be less than $-\sqrt{3}$,
(Hern\'andez-Aguayo \&  Ure\~na-L\'opez 2011). This value for the
$\lambda$ parameter was found using others techniques, in  quantum
solutions and in supersymmetric quantum solutions in quantum
cosmology, for the same flat FRW cosmological model (Socorro \&
D'oleire 2010); (Socorro et al 2013). Chimento \& Jakubi (1996)
 found the value $\lambda= -\sqrt{2}$,  for
inflationary era, solving the Einstein field equations  as power
law.

Is common say that when the scalar potential have a exponential
behavior as in this case, the  universe must have a fast growing
scale factor.
 Remembering that $\alpha_\phi=1+m_\phi$,  and considering that ordinary matter is only
 $4\%$ of the total, and assuming that the scalar field account for the remaining density,
 we need that $\rm \alpha_\phi$ near to 18 when we consider the dark energy scenario.  In that case, the scale factor is fast growing
 and the quintessence field is dominant in the evolution in the
universe.

Comparing the different scale factor in the figure 1, the temporal
potential term to the left graphics, have a big negative slope,
making that the universe roll faster, having a fast growing when the
parameter $\rm m_\phi=18$, in other values this slope is attenuated,
making a moderate expansion in the scale factor.

\begin{figure}[h]
\begin {center}
\includegraphics[width=0.75\textwidth]{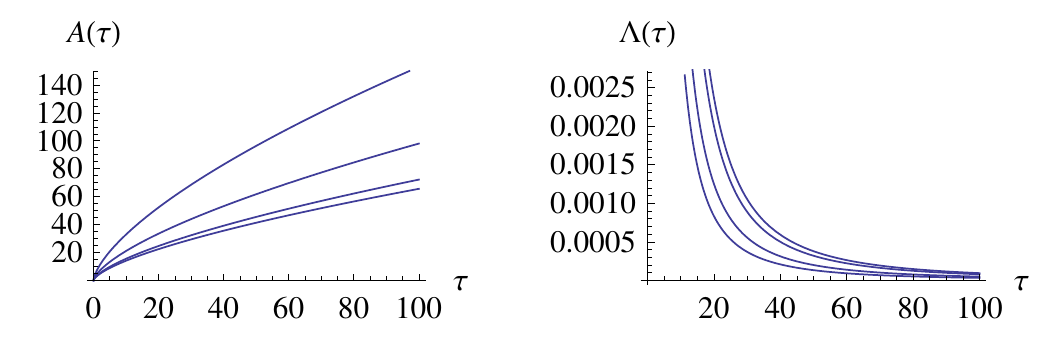}
\caption{In the dust scenario, the scale factor has a fast growing
for big values to the $\rm m_\phi$ parameter, in the plot we choose
the values 0.5,1,4 and 18, going to the down to up side,
respectively, in the graphics. This behavior corresponds to the
cosmological term (potential term), in the right to left side in the
corresponding plot.}
\end{center}
\end{figure}

In particular, in  the temporal potential field there is a different
behavior in the axe $m_\phi$ in the $[0,1]$, region where the
ordinary matter have a dominant behavior; in the other region, the
quintessence field is dominant in the evolution of the universe.

\item{} Radiation, $\omega=\frac{1}{3}$

The set of equations (\ref{potencial},\ref{kinetic},\ref{pot-phi}),
for this value of the barotropic parameter, have the following form
\begin{eqnarray}
\rm V_\omega(\tau)&=&\rm \frac{
m_\phi}{4\alpha_\phi}\frac{1}{\tau^2},
\label{radiation}\\
\rm \Delta \phi(\tau)&=&\rm \sqrt{\frac{m_\phi}{\alpha_\phi}}
\,Ln(\tau),
\label{phi-radiation}\\
\rm V(\phi)&=&\rm \frac{m_\phi}{4\alpha_\phi}\,
e^{-2\sqrt{1+\frac{1}{m_\phi}}\Delta \phi}. \label{pot-rad}
\end{eqnarray}
\end{enumerate}
and the corresponding scale factor
$$\rm  A_{\frac{1}{3}}(\tau)=\left[a_{\frac{1}{3}} \, \tau
\right]^{\frac{1}{2}},\quad a_{\frac{1}{3}}=\frac{4}{3}\sqrt{6\pi G
\alpha_\phi M_{\frac{1}{3}}}.$$

\begin{figure}[ht!]
\begin {center}
\includegraphics[width=0.75\textwidth]{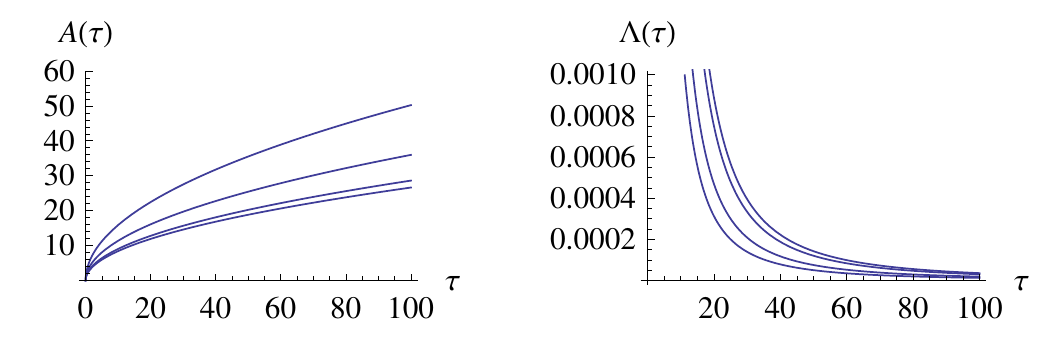}
\caption{In the radiation scenario, the scale factor has a fast
growing for big values to the $\rm m_\phi$ parameter, in the plot we
choose the values 0.5,1,4 and 18, going to the down to up side,
respectively, in the graphics. This behavior corresponds to the
cosmological term (potential term), in the right to left side in the
corresponding plot.}
\end{center}
\end{figure}

The temporal dependence of the cosmological term
 goes as $\rm \frac{1}{\tau^2}$,
a result that is reported in all references that used a proportional relation
between the energy density of the scalar field and the energy
density to the barotropic perfect fluid,  in non covariant theory.

\subsection{Negative branch: $\rm -1<\omega<0. $}
In this case we write the relation between the field pressure and density as
\begin{equation}
p_\phi=-|\omega| \rho_\phi; \to \,
\frac{1}{2}\phi^{\prime2}=\beta_\omega V(\phi),\qquad
\beta_\omega=\frac{1-|\omega_\phi|}{1+|\omega_\phi|}
\end{equation}
and we consider two particular values of $\omega$
\begin{enumerate}
\item{} For instance, when we choose  the case
$\omega_\phi=-\frac{2}{3}$, i.e.,  $\rm |\omega_\phi|=\frac{2}{3}$,

The set of equations (\ref{potencial},\ref{kinetic},\ref{pot-phi})
have the following form
\begin{eqnarray}
\rm V_\omega(\tau)&=&\rm \frac{10m_\phi}{\alpha_\phi}
\frac{1}{\tau^2},
\label{infla1}\\
\rm \Delta \phi(\tau)&=&\rm 2\sqrt{\frac{m_\phi}{\alpha_\phi}} \,
Ln(\tau),
\label{phi-infla1}\\
\rm V(\phi)&=&\rm \frac{10m_\phi}{\alpha_\phi}
e^{-\sqrt{1+\frac{1}{m_\phi}}\Delta \phi} , \label{pot-infla1}
\end{eqnarray}
with the law for the scale factor
\begin{equation}
\rm A_{-\frac{2}{3}}(\tau)=  \left[a_{-\frac{2}{3}}\right]^2 \,
\tau^2, \qquad a_{-\frac{2}{3}}=\frac{1}{3}\sqrt{6\pi G \alpha_\phi
M_{-\frac{2}{3}}}
\end{equation}
We consider that in this phenomenological scenario, the values of
the $\rm m_\phi$ are in the interval  $(0,1]$, for instance when
$\rm m_\phi=1$, we recover the  potential for the scalar  field
given by Chimento \& Jakubi (1996) with the corresponding scale
factor.

\begin{figure}[ht!]
\begin {center}
\includegraphics[width=0.75\textwidth]{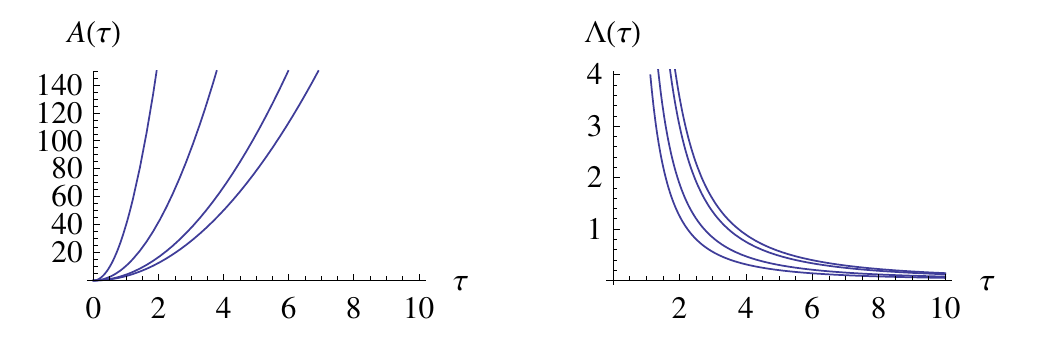}
\caption{In the inflation like scenario, the scale factor has a fast
growing for big values to the $\rm m_\phi$ parameter, in the plot we
choose the values 0.5,1,4 and 18, going to the down to up side,
respectively, in the graphics. This behavior corresponds to the
cosmological term (potential term), in the right to left side in the
corresponding plot.}
\end{center}
\end{figure}

\item{} When we choose $\omega=-\frac{1}{3}$, i.e.,
$\rm |\omega_\phi|=\frac{1}{3}$, we have
\begin{equation}
\rm V(\tau)=\frac{2m_\phi}{\alpha_\phi} \frac{1}{\tau^2},
\end{equation}
so, the scalar field is
\begin{equation}
\rm \Delta \phi=\sqrt{\frac{2m_\phi}{\alpha_\phi}} \, Ln(\tau),
\end{equation}
thus, we can write $\rm V(\phi)$
\begin{equation} \rm
V(\phi)=\frac{2m_\phi}{\alpha_\phi}
e^{-\sqrt{2\left(1+\frac{1}{m_\phi}\right)}\Delta \phi} ,
\end{equation}
with a linear evolution for the scale factor
\begin{equation}
\rm A_{-\frac{1}{3}}(\tau)=  a_{-\frac{1}{3}} \, \tau, \qquad
a_{-\frac{1}{3}}=\frac{2}{3}\sqrt{6\pi G \alpha_\phi
M_{-\frac{1}{3}}}.
\end{equation}
\end{enumerate}

\section{Conclusions}
In this work we have characterized the cosmological term
$\Lambda(\tau)$ as proportional  the potential for the scalar field.
Assuming a proportionality between the energy density of the scalar
field and the the density of a barotropic fluid of the  matter
content, an also assuming that the pressure and density of the
scalar field satisfy a barotropic law, so that  the field equation
in the case of the FRW metric reduces to the standard cosmology in
term of a total energy density and pressure that also satisfy a
barotropic law. We found that for consistency all the different
barotropic parameters should be the same. In the case of flat space
we were able to find general exact solutions. A common
characteristic of all the solutions presented here is that the
dynamic cosmological "constant" is decreasing in time as $\rm
\frac{1}{\tau^2}$ and that it is an exponential function of the
scalar field. We also found the exponential behavior in the scalar
field in the evolution of the universe, and in particular case, the
dust era $\omega_\phi=0$, the scalar potential have a time
dependence in agreement with others results that uses directly
quintessence field in the dark energy frame,  signal that the
universe must have a fast growing  scale factor. This fast growing
scale factor correspond to  $\rm m_\phi
>1$, that is  when the
quintessence field  dominates in the universe, by instance  we
claim that if the percent to usual matter becomes as $4\%$,
in the dark energy and dark matter scenario
 we have  that $\rm \alpha_\phi$ is near to 18. However the case
$\rm m_\phi <1$  corresponds to scaling behavior with the usual
matter. In all epochs analyzed in this work using the relation
between the energy density of the scalar field and the energy
density of the ordinary matter, the behavior of the cosmological
term goes as $\frac{1}{\tau^2}$, these results were found by other
authors in a non covariant way (Chen \& Wu 1990); (Abdel 1990);
(Pavon 1991); (Carvalho et al 1992); (Kalligas et al 1992); (Lima \&
Maia 1994);  (Lima \& Carvalho 1994); (Lima \& Trodden 1996); (Arbab
\&  Abdel 1994); (Birkel \& Sarkar 1997); (Silveira \& Waga 1997);
(Starobinsky 1998); (Overduin \& Cooperstock 1998); (Vishwakarma
2000,2001); (Arbab 2001,2003,2004); (Cunha \& Santos 2004);
(Carneiro \& Lima 2005); (Fomin et al 2005); (Sola \& Stefancic
2005,2006); (Pradhan et al 2007); (Jamil \& Debnath 2011) and
(Mukhopadhyay 2011).
 We consider that this behavior is dependent on the
relation between the energy densities considered in this work and
others.

\acknowledgments{ \noindent This work was partially supported by
CONACYT  167335, 179881 grants. PROMEP grants UGTO-CA-3 and
UAM-I-43. This work is part of the collaboration within the
Instituto Avanzado de Cosmolog\'{\i}a and Red PROMEP: Gravitation
and Mathematical Physics under project {\it Quantum aspects of
gravity in cosmological models, phenomenology and geometry of
space-time}. Many calculations where done by Symbolic Program REDUCE
3.8.}

\vglue 2cm

\noindent REFERENCES\\

\noindent Chen, W., \&  Wu, Y.S., Phys. Rev. D {\bf 41},695(1990).\\
\noindent Abdel-Rahman, A.M.M., Gen. Rel. Grav. {\bf 22}, 655 1990).\\
\noindent Pavon, D., Phys. Rev. D {bf 43}, 375 (1991).\\
\noindent Carvalho, J.C. et al.,  Phys. Rev. D {\bf 46}, 2404 (1992).\\
\noindent  Kalligas, D. et al., Gen. Rel. Grav.{\bf 24}, 351 (1992).\\
\noindent Lima, J.A.S., \&  Maia, J.M.F., Phys. Rev. D {\bf 49}, 5597 (1994). \\
\noindent Lima, J.A.S., \& Carvalho, J.C., Gen. Rel. Grav. {\bf 26}, 909 (1994).\\
\noindent Lima, J.A.S., \& Trodden, M., Phys. Rev. D {\bf 53}, 4280 (1996).\\
\noindent Arbab, A.I., \&  Abdel-Rahaman, A.M.M., Phys. Rev. D {\bf 50}, 7725 (1994).\\
\noindent Birkel, M., \& Sarkar, S., Astropart. Phys. {\bf 6}, 197 (1997).\\
\noindent Silveira V., \& Waga, I., Phys. Rev. D {\bf 56}, 4625 (1997).\\
\noindent Starobinsky, A.A., JETP Letters {\bf 8}, 757 (1998). \\
\noindent J.M. Overduin, J.M., \& Cooperstock, F.I., Phys. Rev. D {\bf 58}, 043506 (1998).\\
\noindent Vishwakarma, R.G., Class Quantum Grav. {\bf 17}, 3833 (2000).\\
\noindent Vishwakarma, R.G., Gen. Relativ. Gravit. {\bf 33}, 1973 (2001).\\
\noindent Arbab, A.I., Spacetime and substance {\bf 1}, 39 (2001).\\
\noindent Arbab, A.I., Class. Quantum Grav. {\bf 20}, 93 (2003).\\
\noindent Arbab, A.I., Astrophys. Space Sci. {\bf 291}, 141 (2004).\\
\noindent Cunha, J.V., \&  Santos, R.C., Int. J. Mod. Phys. D {\bf 13}, 1321 (2004). \\
\noindent Carneiro, S., \&  Lima, J.A.S., Int. J. Mod. Phys. A {\bf 20}, 2465 (2005).\\
\noindent Fomin, P.I., et al., preprint [gr-qc/0509042].\\
\noindent Sola, J., \& Stefancic, H., Phys. Lett. B {\bf 624}, 147 (2005). \\
\noindent Sola, J., \& Stefancic, H., Mod. Phys. Lett. A {\bf 21}, 479 (2006). \\
\noindent Pradhan, A. et al., Rom. J. Phys. {\bf 52}, 445 (2007).\\
\noindent Jamil, M., \& Debnath, U., Int. J. of Theor. Phys. {\bf 50}, 1602 (2011).\\
\noindent Mukhopadhyay, U. et al.,  Int. J. of Theor. Phys. {\bf 50}, 752 (2011). \\
\noindent Aroonkumar Beesham, Phys. Rev. D {\bf 48}, 3539 (1993).\\
\noindent Aroonkumar Beesham, Gen. Rel. Grav. {\bf 26}, 159 (1994).\\
\noindent Arbab, A.I., Gen. Rel. Grav. {\bf 29}, 61 (1997).\\
\noindent Singh, T. et al., Gen. Rel. Grav. {\bf 30}, 573 (1998). \\
\noindent Pradhan, A. \&  Kumar, A., Int. J. of Mod. Phys. D {\bf 10}, 291 (2001).\\
\noindent Pradhan, A., Int. J. of Mod. Phys. D {\bf 12}, 941 (2003).\\
\noindent Pradhan, A., Fizika B {\bf 16}, 205 (2007). \\
\noindent Pradhan, A., Commun. Theor. Phys. {\bf 51}, 367 (2009).\\
\noindent Pradhan, A., \& Pandey, A.P., Int. J. of Mod. Phys. D {\bf 12}, 1299 (2003).\\
\noindent Pradhan, A., \&  Pandey, A.P., Astrophys. and Spa. Sci. {\bf 301}, 127 (2006).\\
\noindent Pradhan, A. et al., Int. J. of Theor. Phys. {\bf 46}, 2774 (2007).\\
\noindent Pradhan, A. et al., Brazilian J. of Phys. {\bf 38}, 167 (2008). \\
\noindent Pradhan, A. et al.,  Astrophys. Space Sci. {\bf 337}, 401 (2012).\\
\noindent Carneiro, S., Int. J. of Mod. Phys. A {\bf 20}, 2465 (2005).\\
\noindent Esposito, G., et al., Class. Quantum Grav. {\bf 24}, 6255 (2007).\\
\noindent Bal, R., \& Singh, J.P., Int. J. of Theor. Phys. {\bf 47}, 3288 (2008).\\
\noindent Belinch\'on, J.A., Int. J. Mod. Phys. A {\bf 23}, 5021 (2008).\\
\noindent Singh, J.P. et al., Astrophys. Spa. Sci. {\bf 314}, 83 (2008).\\
\noindent Singh, M.K. et al., Int. J. of Phys. {\bf 1}, 77 (2013).\\
\noindent Shen, M., Int. J. of Theor. Phys. {\bf 52}, 178 (2013).\\
\noindent Tripathy, S.K., Int. J. Theor. Phys. {\bf 52}, 4218 (2013).\\
\noindent Rahman, M.A., \& Ansary, M.,Prespace time J. {\bf 4}, 871 (2013). \\
\noindent Ferreira, P.G., \& Joyce, M., Phys. Rev. Lett. {\bf 79}, 4740 (1997).\\
\noindent Ferreira, P.G., \& Joyce, M., Phys. Rev. D {\bf 58}, 023503 (1998).\\
\noindent Liddle, A.R., \& Sharrer, R.J., Phys. Rev. D {\bf 59}, 023509 (1998).\\
\noindent Copeland, E.J. et al.,  Phys. Rev. D {\bf 57}, 4686 (1998).\\
\noindent Chimento, L.P., \& Jakubi, A.S., Int. J. of Mod. Phys D {\bf 5}, 71 (1996).\\
\noindent Reyes, M.A., preprint [arXiv:0806.2292].\\
\noindent Lucchin, F., \& Matarrese, S., Phys. Rev. d {\bf 32}, 1316 (1985).\\
\noindent Halliwell, J., Phys. Lett. B {\bf 185}, 341 (1985).\\
\noindent Burd, A.B., \& Barrow, J.D., Nucl. Phys. B {\bf 308}, 929 (1988).\\
\noindent Weetterich, C., Nucl. Phys. B {\bf 302}, 668 (1998).\\
\noindent Wand, D. et al., Ann (NY) Acad. Sci. {\bf 688}, 647 (1993).\\
\noindent Berbena, S.R. et al., Rev. Mex. F\'is. S {\bf 53} (2) 115 (2007).\\
\noindent Riess, A.G., et al., astron. J. {\bf 116}, 1009 (1998).\\
\noindent Perlmutter, S. et al.,  astrophys. J. {\bf 517}, 565 (1999). \\
\noindent Sabido, M. et al., Fizika B {\bf 19} (4), 177-186 (2010).\\
\noindent Socorro, J. et al., Advances in High Energy Phys. 805164 (11p) (2014).\\
\noindent Hern\'andez-Aguayo, C., \&  Ure\~na-L\'opez, L.A., AIP Conf. Proc. {\bf 1473}, 68 (2011).\\
\noindent Socorro, J., \&  D'oleire, M., Phys. Rev. D. {\bf 82}(4) 044008 (2010).\\
\noindent Socorro, J. et al., Cap\'itulo cuatro, book: Procesos no
    lineales en la ciencia y la sociedad,
    (2013), Edit. Notabilis Scientia, Compilador M. A. Granados; ISBN: 978-607-8289-41-7, in {\it
    Inflaci\'on Cosmol\'ogica Vista desde la Mec\'anica Cu\'antica
    Supersim\'etrica}.\\

\end{document}